\documentclass[twocolumn,prl]{revtex4}
\usepackage{graphicx}
\usepackage{dcolumn}
\usepackage{bm}
\usepackage{hyperref}
\usepackage{amsmath}

\begin{document}

\author{Edwin Barnes$^{1}$ and Sophia E. Economou$^2$}
\affiliation{ $^{1}$Condensed Matter Theory Center, Department of
Physics, University of Maryland, College Park, Maryland 20742-4111, USA\\
$^{2}$Naval Research Laboratory, Washington, DC 20375, USA}

\title{Electron-nuclear dynamics in a quantum dot under non-unitary electron control}

\begin{abstract}
We introduce a method for solving the problem of an externally
controlled electron spin in a quantum dot interacting with host
nuclei via the hyperfine interaction. Our method accounts for
generalized (non-unitary) evolution effected by external controls
and the environment, such as coherent lasers combined with
spontaneous emission. As a concrete example, we develop the
microscopic theory of the dynamics of nuclear-induced frequency
focusing as first measured in Science \textbf{317}, 1896 (2007); we
find that the nuclear relaxation rates are several orders of
magnitude faster than those quoted in that work.
\end{abstract}

\maketitle

The nuclear environment in III-V quantum dots has been recognized in
recent years as the main source of decoherence for the electron spin
and thus constitutes an important hurdle for quantum technologies
with these systems. The microscopic dynamics of the closed
electron-nuclear spin system have been investigated in important
recent theoretical contributions
\cite{wangluke,coishschliemanndenghu}. In these works
\cite{wangluke}, controls have been represented as ideal, unitary
rotations of the electron spin, and the nuclear polarization along the external field is
taken to be unaltered during the electron evolution. This no longer
is the case in experiments involving controls that couple the system
to an additional bath, which can exchange polarization with the
system. Such experiments are relevant because incoherent
interactions are needed to initialize and read out the system. These
experiments in quantum dots (QDs) observed dynamic nuclear
polarization and nuclear feedback effects
\cite{greilichnuclei,bracker,sam}. While the details of the various
experiments differ, the main common feature is that an external
control interacts with the electron, and through the hyperfine
interaction the nuclear spins are also partially polarized. The
theories employed to describe such experiments are usually in the
form of rate equations and some sort of Fermi golden rule, and
typically invoke phenomenological terms. Other theories \cite{christtaylor} employed a more microscopic approach, but without including, e.g., feedback and a complete treatment of control fields.

In this Letter, we develop a theory that addresses such experiments
involving non-unitary evolution of the electron while
still treating the electron-nuclear interaction microscopically. We
make use of the operator sum representation of quantum evolution and
its simplified form in the spin vector (SV) representation and
develop a theory that is perturbative with respect to the hyperfine
coupling. We develop both Markovian and non-Markovian treatments,
and by comparison of the two we establish the validity regime of the Markovian approximation.

In order to illustrate the power of our approach,
we apply it to the experiment of Ref.\cite{greilichnuclei}, a proper
microscopic theory of which is lacking to date. This experiment
demonstrated nuclear-induced focusing of the electron precession
rates in a QD ensemble through the feedback dynamics of the electron
and nuclear spins. This mechanism is largely driven by non-unitary
evolution of the electron spin, making it difficult to solve
conventional Master Equations to analyze the dynamics. Instead, a
phenomenological treatment was introduced in the Supporting Online
Material of \cite{greilichnuclei} and further developed in \cite{sam}. Our microscopic solution does
not invoke phenomenological quantities and provides a unified
description of the experiments in \cite{greilichnuclei,sam}. One of
our striking results is that the nuclear relaxation process is
several orders of magnitude faster than what is used in \cite{greilichnuclei,sam}.

The system we consider is a single electron trapped in a QD and
subject to an external in-plane static magnetic field $B_z$, which
splits the spin states along the $z$ direction. The electron
interacts through the hyperfine contact interaction with $N$ nuclear
spins in the QD ($N\approx10^5$). There is also an external
time dependent field acting on the electron, as well as the photon
bath that drives spontaneous emission (electron-hole recombination).
The total Hamiltonian is $H=H_0+H_{hf}+H_p+H_{rad}$, where
\begin{eqnarray}
H_0 &=& \omega_e \hat{S}_z + \epsilon_T |T\rangle\langle T| +
\omega_n {\sum}_i \hat{I}^i_z
\\
H_{hf} &=& {\sum}_i A_i \hat{S}_z \hat{I}^i_z + {\sum}_i
{A_i}/{2}(\hat{S}_+\hat{I}^i_-+\hat{S}_-\hat{I}^i_+)
\\
H_p &=& \Omega(t) |\bar{x}\rangle\langle T| + h.c.
\\
H_{rad} &=&  {\sum}_kg_k (|z\rangle\langle T|+|\bar z\rangle\langle
T|) a_k^\dagger e^{i\omega_k t} + h.c.
\end{eqnarray}
In Eqs. (1)-(4), $\hat{S}_j(\hat{I}_j)$ is the electron
(nuclear) spin operator along the $j$ axis,
$\hat{S}_\pm=\hat{S}_x\pm i\hat{S}_y$, $\Omega(t)$ contains the
pulse information, $|z\rangle(|\bar{z}\rangle)$ is the spin up (down) state along the B-field
direction,  $|\bar{x}\rangle$ is the spin down state along
the optical axis $x$, $|T\rangle$ is the excited trion
state, $g$ is the coupling to the radiation bath, and $a^\dagger$ is
the bath photon creation operator. In the hyperfine
Hamiltonian $H_{hf}$, the first term is referred to as the Overhauser
term, while the second is called the `flip-flop' term.

The couplings $\Omega(t)$ arise from periodic, ultrafast laser
fields like those used in \cite{greilichnuclei}. Since
$A_i/\omega_e\ll 1$ for moderate magnetic fields, we give a
perturbative treatment in this small parameter. We first focus on
the zeroth order solution (electron spin periodically driven without
nuclear coupling).
\begin{figure}[t]
\begin{center}
\includegraphics[width=8 cm,clip=true, bb=6.6cm 4.6cm 24.5cm 12.5cm]{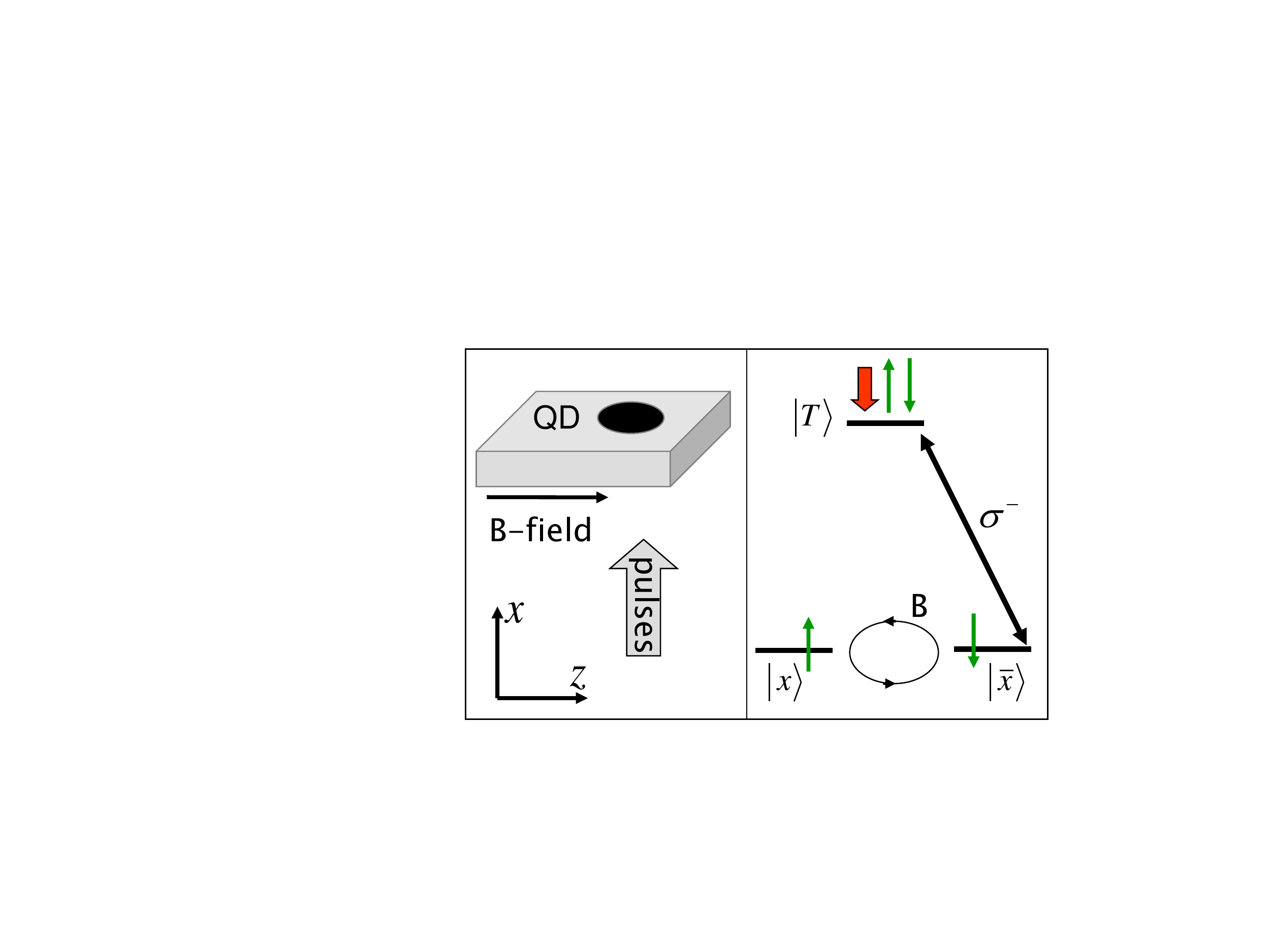}
\caption{Color online. Left: Experimental setup of \cite{greilichnuclei,sam}. Right: Relevant QD states and polarization selection rules.}\label{levels}
\end{center}
\end{figure}
The primary effect of the pulses on the electron spin dynamics is
the creation or destruction of spin polarization depending on the
spin state. This arises from the selection rules of the three-level
system in conjunction with the perpendicular external magnetic
field. For concreteness, we consider $\sigma^-$ pulses, in which
case only $|\bar x\rangle$ is coupled to
the light and excited by it to the trion, see Fig. \ref{levels}.
Depending on the pulse parameters, a certain population is moved to
the trion. This population subsequently decays back to
the spin subspace via spontaneous emission of a photon. Due to the
B-field, the population decays equally to the $|x\rangle$ and $|\bar x\rangle$ states, changing the
electron spin polarization \cite{economou05}.

This physics describes non unitary evolution of the electron spin
due to the coupling of the system to the photon bath. To describe
this mathematically \emph{in the spin subspace} we need a
generalization of the usual unitary evolution operator to a set of
so-called Kraus operators $\{E_j\}$ which transform the density
matrix as $\sum_j E_j \rho E^\dagger _j$ \cite{nielsenchuang}. These
can be found by solving for the non-unitary part of the evolution of
an arbitrary initial system density matrix and relating it to the
final density matrix. Following this standard procedure \cite{supplement} we
find the following Kraus operators in the $|x\rangle, |\bar
x\rangle$ basis
\begin{eqnarray}
E_1=\left[\begin{array}{cc}
1 & 0 \\
0 & q \\
\end{array}
\right],
~
E_2=\left[\begin{array}{cc}
0 & k \\
0 & 0 \\
\end{array}
\right],
~
E_3=\left[\begin{array}{cc}
0 & 0 \\
0 & k \\
\end{array}
\right],
\label{kraus}
\end{eqnarray}
where $q=q_o e^{i \phi}$ and $k=\sqrt{\frac{1-q_o^2}{2}}$. The parameter $q_o^2$
is the probability to go from $|\bar{x}\rangle$ to
$|\bar{x}\rangle$; it is related to the pulse area and takes values
from 0 to 1. The quantity $(1-q_o^2)$ is related to the probability
of population remaining in the trion state after the passage of the
pulse, and thus $q_o$ quantifies the deviation from unitary dynamics
in the qubit subspace (for unitary evolution $q_o=1$). The parameter
$\phi$ is the spin rotation angle caused by the pulse and is a
function of the detuning. We have therefore found 2-d matrices to
describe the more complicated dynamics of the pulse followed by
spontaneous emission.

In between pulses the evolution is simply Larmor precession under
$B_z$, given by $U=e^{-i\omega_e T_R\hat{S}_z}$, where $ T_R$ is the
period of the pulse train. We are interested in finding the steady
state electron spin. For this, the SV representation ($S_{e,j}=2\hbox{Tr}(\rho\hat{S}_j)$) is most
convenient as all the operations act on the left side of the SV. As
a result of the non-unitarity of the evolution, in addition to the
transformation of the SV a new contribution is generated at each
cycle:
\begin{eqnarray}
S_e(n T_R) = S_e^{(n)} = Y_e S_e^{(n-1)} + K_e, \label{se3}
\end{eqnarray}
where we found that $(Y_e)_{ij}=2\sum_\ell \text{Tr}\left[\hat{S}_i
E_\ell U \hat{S}_j U^\dagger E_\ell^\dagger\right],$
$(K_e)_j=2\sum_\ell\text{Tr} \left[\hat{S}_j E_\ell
E_\ell^\dagger\right]$. In the limit $n\rightarrow\infty$ the steady
state is $S_e^{(\infty)} = (1-Y_e)^{-1} K_e$ (explicit expression is
in \cite{supplement}). We therefore see that a $3\times 3$ matrix,
$Y_e$, and a three-dimensional vector, $K_e$, are the quantities
that determine the dynamics of the electron spin. Because its
structure is convenient we use the equivalent and more compact
$4\times 4$ matrix that contains all the information:
\begin{eqnarray}
\mathcal{Y}_e=\left[
\begin{array}{cccc}
1 & 0 & 0 & 0 \\
K_{e,x} & Y_{e,xx} & Y_{e,xy} & Y_{e,xz} \\
K_{e,y} & Y_{e,yx} & Y_{e,yy} & Y_{e,yz} \\
K_{e,z} & Y_{e,zx} & Y_{e,zy} & Y_{e,zz}
\end{array}%
\right]. \label{espin}
\end{eqnarray}
In this 4-d representation, the steady-state SV
$\mathcal{S}_e^{(\infty)}=(1,S_{e,x}^{(\infty)},S_{e,y}^{(\infty)},S_{e,z}^{(\infty)})
$ is the eigenvector of $1-\mathcal{Y}_e$ with eigenvalue 0. This
more compact representation will prove very useful when we
introduce the nuclear spin. 

Having solved the zeroth order problem, we proceed to the inclusion
of the hyperfine interaction. For simplicity we assume that the
nuclear spin has $I=1/2$. The nuclear spins affect each other
through their interactions with the electron. When $A\sqrt{N}/\omega_e\ll
1$, where $A$ is a typical value of $A_i$, flip-flops occur slowly so that multinuclear effects such
as dark state saturation \cite{christtaylor} are negligible, and the primary effect of the nuclear spins on the electron is a
shift of the precession frequency through the Overhauser term
(Overhauser shift). Therefore, we consider first a single nuclear
spin interacting with the electron and incorporate multinuclear
effects by shifting the electron Zeeman frequency by an amount
proportional to the net nuclear polarization
\cite{greilichnuclei,sam}.

For a single nuclear spin interacting with the electron spin via the
hyperfine Hamiltonian we use a SV representation, which in this case
is 15-d. For the type of control used in Ref.
\cite{greilichnuclei,sam}, there are no nuclear effects during the
ultrashort (i.e., broadband) pulses, which do not distinguish
between the electron spin eigenstates along the field $B_z$.
Therefore, the Kraus operators are simply tensor products between
the $E_j$'s of Eq. (\ref{kraus}) and the identity. Following the same prescription as for the single spin, we define a
16-d SV $ \mathcal{S}_i= 4\text{Tr}\left(\rho G_i \right),$ where
$\rho$ is the $4\times4$ density matrix of the two spins and the
generators $G_i$ are tensor products of spin operators (including
the identity) $G_{4k+\ell}=\hat{S}_k\otimes\hat{I}_\ell$, where
$k,\ell$ run from 0 to 3. With our conventions, $\mathcal{S}_0=1$.
The 16-d analogue of $\mathcal{Y}_e$ is given by
$\mathcal{Y}_{ij}=4\sum_\ell \text{Tr}\left[G_i E_\ell U_{e,n} G_j
U_{e,n}^\dagger E_\ell^\dagger\right]$. In general $\mathcal S$ is not
simply a tensor product of the two individual SVs, but contains
quantum correlations (entanglement).

The pulses are expected to `interrupt' the electron-nuclear evolution
only for $q_o \ll 1$, while entanglement will build up when $q_o\sim
1$. Therefore a Markovian approximation should be sufficient for short pulse train
periods and pulses of strength $q_o\sim 0$ (as in \cite{greilichnuclei,sam}) . 

\emph{Markovian approximation}--To find an effective relaxation rate
for the nuclear spin, we use the equation $ \mathcal{S}(t+T_R) =
\mathcal{Y} \mathcal{S}(t) $ for the 16-d case. In the Markovian
approximation, we only keep the separable (tensor product) part of
$\mathcal{S}$, i.e.,
$\mathcal{S}=\mathcal{S}^{(\infty)}_e\otimes\mathcal{S}_n$, where we
have used that the timescales of evolution for the electron
and the nuclei are quite different \cite{merkulov}, so that we can
assume that the electron steady state
is reached quickly compared to the
nuclear dynamics \cite{greilichnuclei,sam,nazarov,rudnerlevitov}. The equation for the 4-d nuclear SV is then
$\mathcal{S}_n(t+T_R) = \mathcal{Y}_n \mathcal{S}_n(t),$ where
$\mathcal{Y}_n$ explicitly contains electron SV components. Since the nuclear evolution is much slower than the pulse repetition rate, we can coarse grain this equation, and obtain a
differential equation for the nuclear SV, $ \frac{d}{dt}
\mathcal{S}_n = {1\over T_R}(\mathcal{Y}_n-1)\mathcal{S}_n,$ which
gives $\mathcal{S}_n(t)=e^{(\mathcal{Y}_n-1)
t/T_R}\mathcal{S}_n(0).$ For small flip-flop coupling (but keeping
the Overhauser term to all orders), we find the two smallest
eigenvalues of $1-\mathcal{Y}_n$ to be $\lambda_1=0$  and
\begin{eqnarray}
\lambda_2= \frac{A^2}{\omega_e^2}
\frac{1+S_{e,z}^2+(S_e^2-1)\cos(\frac{A
T_R}{2})}{1+S_{e,z}^2+(S_{e,z}^2-1)\cos(\frac{A
T_R}{2})}\sin^2\frac{\omega_e T_R}{2}, \label{lambda2}
\end{eqnarray}
where $S_e$ is the length of the electron steady state SV, and for
brevity we have suppressed the superscript $\infty$ \cite{foot1b}.
The zero eigenvalue corresponds to the nuclear steady-state SV,
which to leading order in the flip-flop term is
$\mathcal{S}_n^{(\infty)} = (1,0,0,S_{n,z}^{(\infty)} )$ where
\begin{eqnarray}
S_{n,z}^{(\infty)}=\frac{2S_{e,z}[\sin^2\left(\frac{A
T_R}{4}\right)+S_{e}^2\cos^2\left(\frac{A
T_R}{4}\right)]}{1+S_{e,z}^2+(S_e^2-1)\cos(\frac{A T_R}{2})}.
\label{svnss}
\end{eqnarray}
\begin{figure}
\begin{center}
\includegraphics[width=7cm,clip=true]{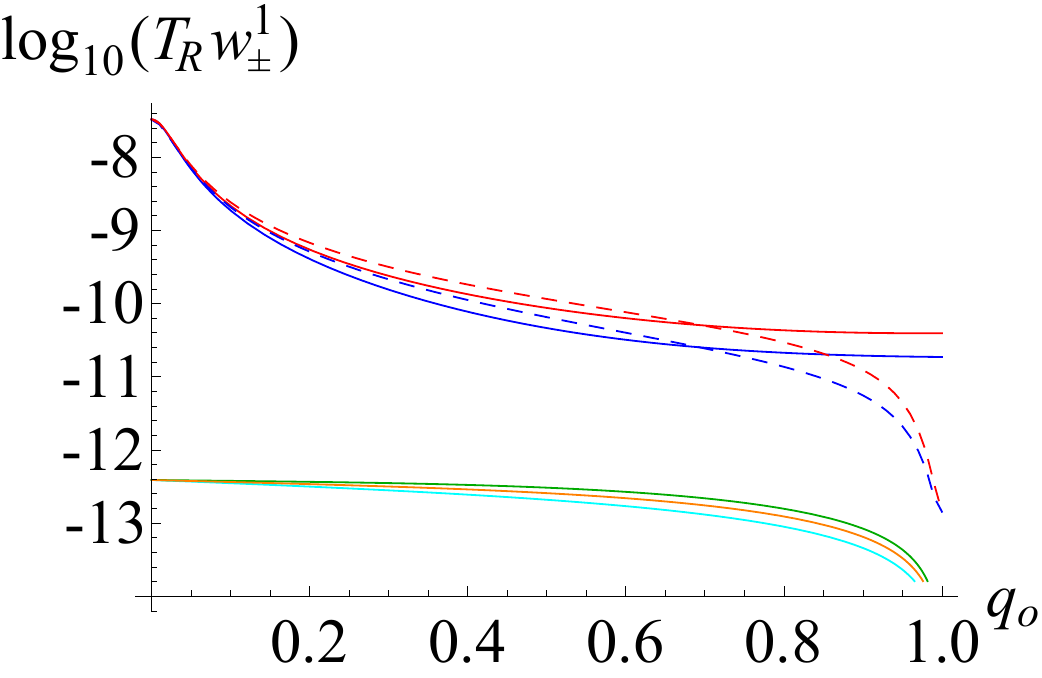}
\caption{Color online. Plot of the log of the nuclear relaxation
rates multiplied by $T_R$ from our current theory (upper set of
curves) with our Markovian approximation (solid) and exact numerical
(dashed) as function of $q_o$. The lower curves are based on Refs.
\cite{greilichnuclei,sam} and are several orders of magnitude less.
Pulse parameters are $\phi=\pi/2,T_R=3900.3/\omega_e,
A/\omega_e=10^{-5}.$}\label{figRelrates}
\end{center}
\end{figure}
The nonzero eigenvalue $\lambda_2$ gives the nuclear relaxation rate
$\gamma_n=\lambda_2/T_R$. The single
nucleus spin flip rates, which are generally different in the
presence of nonzero polarization \cite{sam}, are
$w^{1}_\pm=\gamma_n(1\pm S_{n,z}^{(\infty)})/2$, where $w^{1}_+$
($w^{1}_-$) is the rate to flip from down (up) to up (down). Fig.
\ref{figRelrates} shows that our rates are orders of magnitude
larger than those of \cite{greilichnuclei,sam}. The heuristic
expressions only took into account that the relaxation rates should
vanish when $T_R$ is a multiple of the electron spin precession
period as well as the overall scale factor $A^2/\omega_e^2$. The
first of these features arises because an electron spin synchronized
with the pulses is unaffected by them so that no nuclear relaxation
takes place. The scale factor is fixed by noting that energy
conservation leads to a suppression of hyperfine flip-flops when
$\omega_e\gg\omega_n$; only virtual flip-flops are allowed and since
these must come in pairs, their effect is second-order in
$A/\omega_e$.

Our theory reveals an additional dependence of the relaxation rate $\gamma_n$ on
the orientation of the electron spin that was overlooked
by \cite{greilichnuclei,sam}. When the electron SV is transverse to the B-field
($S_{e,z}\approx0$ and $S_e\gg0$), flip-flops
are not suppressed by energy conservation and angular momentum is freely
transferred from the electron to the nuclei, leading to a strong enhancement of
$\gamma_n$. This is also clear from Eq. (\ref{lambda2}) where the denominator is close
to zero when $S_{e,z}\approx0$ while the numerator remains finite due to $S_e\gg0$.
These conditions are realized in the regime most
relevant for the experiments in Refs. \cite{greilichnuclei,sam} where $q_o\ll1$, that
is, when the pulses drive most of the population out of
the qubit subspace, re-orienting the electron spin along the
optical ($x$) axis. Note that the enhancement of $\gamma_n$ depends crucially on the openness
of the system since the photon bath acts as an angular momentum reservoir. Our predicted timescale can
be checked experimentally by
measuring in a single QD the frequency of the pump-probe signal at various timescales.
By systematically varying the B-field and pulse parameters the
relaxation rates could be mapped out as a function of the
parameters.

\begin{figure}
\begin{center}
\includegraphics[width=8cm,clip=true, bb=1.2cm 5.5cm 25cm 14.2cm]{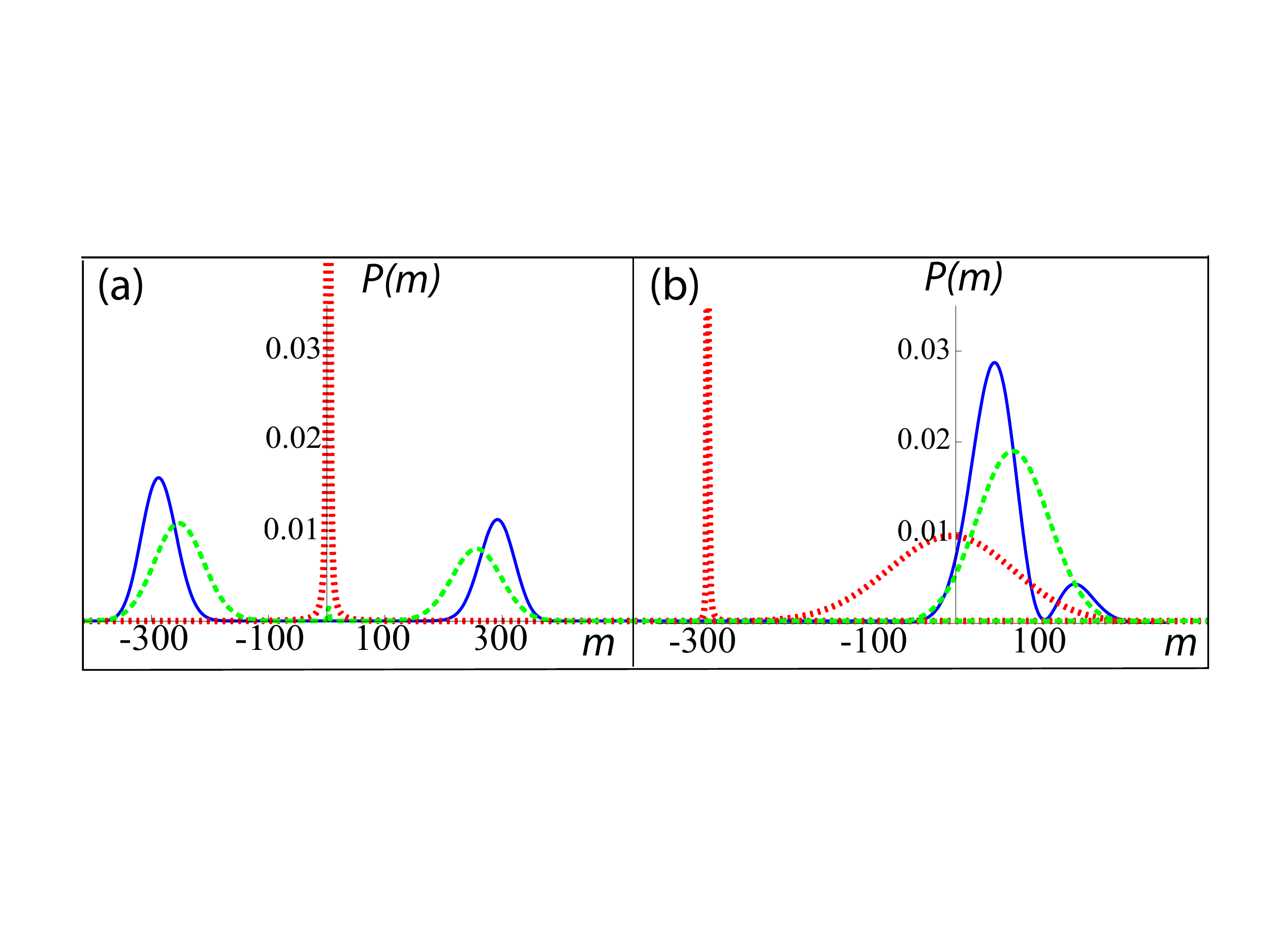}
\caption{Color online. The nuclear polarization probability
distribution with $N=5000$ nuclei and pulse parameters $\phi=\pi/4,
q_o=0.2, A/\omega_e=10^{-5}$ and for (a) $T_R=797.9/\omega_e$ and
(b) $T_R=800.3/\omega_e$, from our Markovian theory (blue/solid
line), the theory from Ref.\cite{greilichnuclei} (red/dotted) and
the one from Ref.\cite{sam}(green/dashed).} \label{figPndist}
\end{center}
\end{figure}

The probability distribution for the net multinuclear
polarization $m/2$ is obtained from a kinetic equation for $m$,
which is the difference in the number of spins pointing up and down:
\begin{eqnarray}
\frac{dP(m)}{dt}&=&-\sum_{\pm}\left[w_\pm(m){N\mp
m\over2}\right]P(m) \label{nucbathRE}
\\
&+&\sum_{\pm}P(m\pm2)w_\mp(m\pm2)\left[{N\pm
m\over2}+1\right],\nonumber
\end{eqnarray}
where $w_\pm(m)$ are the rates in the presence of nuclear
polarization $m/2$. These are found by implementing the Overhauser
shift, $w_\pm(m)=w_\pm^1(\omega_e+ m A)$, where we have assumed
equal couplings for all nuclear spins \cite{foot3}. Examples of the
resulting distribution are shown in Fig. \ref{figPndist} for typical
values of the parameters. In general, large $T_R$ results
in more peaks in $P(m)$ and thus gives rise to a greater degree of
nuclear state ``narrowing" ($T_2^*$ is enhanced). Furthermore, the sharpest peaks occur at
values of $m$ such that $(\omega_e+mA)T_R$ is an odd integer
multiple of $\pi$, and the locations of these peaks can be
controlled by adjusting $\omega_e T_R$. A systematic
exploration of the parameter space can help tailor the nuclear
state.
\begin{figure}
\begin{center}
\includegraphics[width=6cm,clip=true]{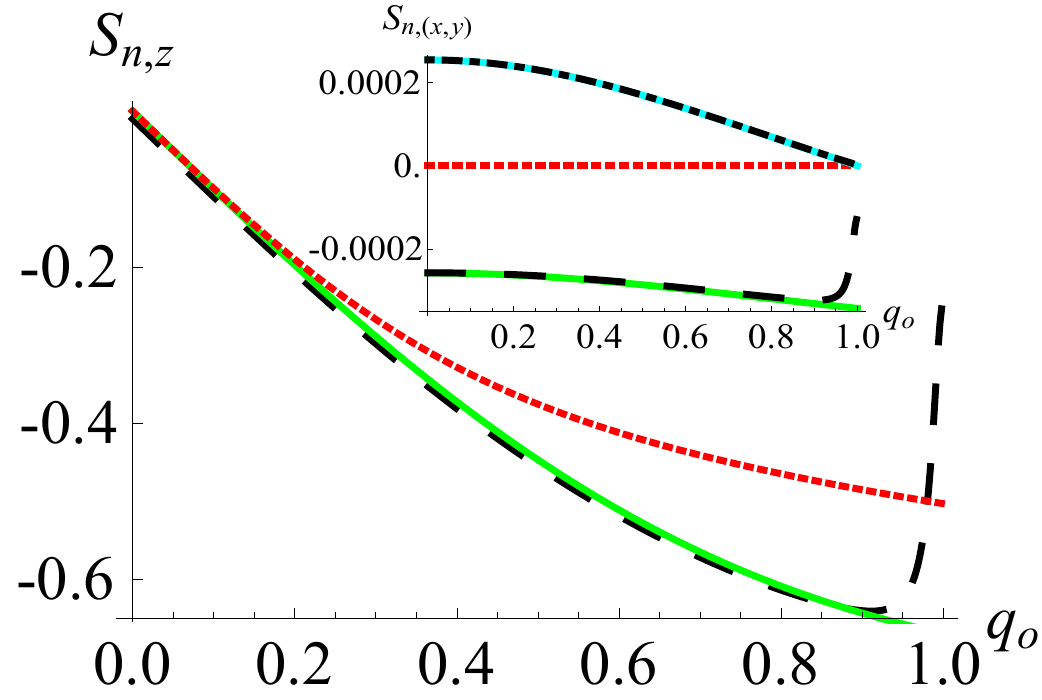}
\caption{Color online. Nuclear steady state SV $z$ component as
function of $q_o$ for Markovian approximation Eq. (\ref{svnss})
(red/dotted), analytical non-Markovian, Eq. (\ref{SSnNonmarkovian})
(green/solid) and exact numerical (black/dashed) for
$\phi=\pi/2,T_R=3900.3/\omega_e,A/\omega_e=10^{-5}$. Inset shows $S_{n,x}$, $S_{n,y}$.} \label{figSSn}
\end{center}
\end{figure}

\emph{Beyond the Markovian approximation}--Our analysis above
provides analytic expressions for the nuclear dynamics in the
Markovian approximation, an approach valid for $q_o\ll1$ (see Fig.
\ref{figRelrates}). Our formalism however is not inherently
Markovian, and we now present an analytical non-Markovian expression
for the nuclear steady state. We return to the 16-d matrix
$\mathcal{Y}$ and perform a perturbative expansion in the coupling
which is a controlled approximation in the hyperfine coupling. The steady state nuclear SV
turns out to be
\begin{eqnarray}
&&S_n^{(\infty)} = c(1+q_o^2-2q_o\cos\phi,(1-q_o^2)\tan(\omega_e
T_R/2),\nonumber\\&&\qquad\qquad q_o \omega_e T_R \sin\phi),
\label{SSnNonmarkovian}
\end{eqnarray}
where $c$ is in \cite{foot2}. Nonzero $x,y$
components arise from expanding the Overhauser
interaction in addition to the flip-flop in deriving Eqn.
(\ref{SSnNonmarkovian}). Fig. \ref{figSSn} shows that the dynamics become less Markovian as the pulses
become more unitary ($q_o\rightarrow 1$).

In conclusion, we have developed a formalism for analyzing
experiments with generalized, non-unitary controls on the electron
spin confined in a QD and coupled to the host nuclei. By applying it
to the experiments of \cite{greilichnuclei,sam} we have found that
the nuclear relaxation is orders of magnitude faster than previously
thought. Our method is in general non-Markovian and is applicable to
controls other than ultrafast lasers by appropriate choice of the
Kraus operators. It can have wide application to other systems, such
as gated QDs and NV centers in diamond \cite{dutt_nv}. An
interesting application of the theory would be to use it for the
design of the final nuclear state.

This work was supported by LPS/NSA (EB) and in part by ONR and LPS/NSA
(SEE).


\begin{thebibliography}{6}
\expandafter\ifx\csname natexlab\endcsname\relax\def\natexlab#1{#1}\fi
\expandafter\ifx\csname bibnamefont\endcsname\relax
 \def\bibnamefont#1{#1}\fi
\expandafter\ifx\csname bibfnamefont\endcsname\relax
  \def\bibfnamefont#1{#1}\fi
\expandafter\ifx\csname citenamefont\endcsname\relax
  \def\citenamefont#1{#1}\fi
\expandafter\ifx\csname url\endcsname\relax
  \def\url#1{\texttt{#1}}\fi
\expandafter\ifx\csname urlprefix\endcsname\relax\def\urlprefix{URL }\fi
\providecommand{\bibinfo}[2]{#2}
\providecommand{\eprint}[2][]{\url{#2}}



\bibitem{wangluke} W. Yao, R.-B. Liu, and L. J. Sham, Phys. Rev. B \textbf{74}, 195301 (2006); L. Cywinski, W. M. Witzel
and S. Das Sarma,  Phys. Rev. B \textbf{79}, 245314 (2009).

\bibitem{coishschliemanndenghu} W. A. Coish, J. Fischer and D. Loss,  Phys. Rev. B \textbf{81}, 165315 (2010); B. Erbe and J. Schliemann,
Phys. Rev. Lett. \textbf{105}, 177602 (2010); C Deng and X. Hu, Phys. Rev. B \textbf{73} 241303(R)
(2006).

\bibitem{greilichnuclei} A. Greilich \textit{et al.}, Science \textbf{317}, 1896 (2007).


\bibitem{bracker} A. S. Bracker \textit{et al.}, Phys. Rev. Lett. \textbf{94}, 047402 (2005);
D. J. Reilly \textit{et al.}, Science \textbf{321}, 817 (2008); X. Xu \textit{et al.}, Nature \textbf{459}, 1105 (2009);
Foletti \textit{et al.}, Nature Phys. \textbf{5}, 903 (2009); C. Latta \textit{et al.}, Nature Phys. \textbf{5}, 758 (2009);
T. D. Ladd \textit{et al.}, Phys. Rev. Lett. \textbf{105}, 107401 (2010); E. A. Chekhovich \textit{et al.}, Phys. Rev. Lett. \textbf{104}, 066804 (2010).


\bibitem{sam} S. G. Carter \textit{et al.}, Phys. Rev. Lett. \textbf{102}, 167403 (2009).

\bibitem{christtaylor} H. Christ \textit{et al.}, Phys. Rev. B \textbf{75}, 155324 (2007); J. Taylor \textit{et al.}, Phys. Rev. Lett. \textbf{91}, 246802 (2003).

\bibitem{economou05} S. E. Economou \textit{et al.}, Phys. Rev. B \textbf{71}, 195327
(2005).

\bibitem{nielsenchuang}  M. A. Nielsen and I. L. Chuang, Quantum Computation and Quantum Information (Cambridge
Univ.
Press).

\bibitem{supplement} See supplement.

\bibitem{merkulov} I. A. Merkulov, Al. L. Efros, and M. Rosen, Phys. Rev. B \textbf{65}, 205309
(2002).

\bibitem{nazarov} J. Danon and Yu. V. Nazarov,  arXiv:1011.3378 (2010).

\bibitem{rudnerlevitov} M. S. Rudner and L. S. Levitov, Phys. Rev. Lett. \textbf{99}, 246602
(2007).

\bibitem{foot1b}   We drop the nuclear Zeeman splitting, since $\omega_n\ll\omega_e$.

\bibitem{foot3} The case of unequal hyperfine couplings can be treated by introducing a probability distribution $P_i(m)$
and flip rates $w^{(i)}_\pm(m)$ for each different value of the coupling $A_i$ and solving a set of rate equations like Eq. (\ref{nucbathRE}).

\bibitem{foot2} $c = \frac{\omega_e T_R \sin{(\omega_e T_R)}}{\xi}
\big[1+q_o^2-2q_o\cos\phi\cos^2(\omega_e T_R/2)\big]
\times\big\{(1+q_o^2)(1+\omega_e^2 T_R^2/2-\cos(\omega_e
T_R))-2q_o(1+(\omega^2_e T^2_R/2-1)\cos(\omega_e
T_R))\cos\phi\big\}^{-1}$, where $\xi$ is in \cite{supplement}.

\bibitem{dutt_nv} M. V. G. Dutt \textit{et al.}, Science \textbf{316}, 1312 (2007).

\end{thebibliography}
\end{document}